\documentclass{elsart}

\usepackage{graphicx}
\usepackage{amsmath}
\usepackage{epstopdf}

\usepackage{epsfig} 

\usepackage{amssymb}

\newcommand{\solarmass}{M_{\odot}}

\def\3EG{{3EG J1746-2851}}
\def\p0{{$\pi^0$}}
\def\1018{{$10^{18}$}}

\newcommand{\be}{\begin{equation}}
\newcommand{\ee}{\end{equation}}
\newcommand{\bea}{\begin{eqnarray}}
\newcommand{\eea}{\end{eqnarray}}
\newcommand{\bmu}{\begin{multline}}
\newcommand{\emu}{\end{multline}}

\def\simlt{\lower.5ex\hbox{$\; \buildrel < \over \sim \;$}}
\def\simgt{\lower.5ex\hbox{$\; \buildrel > \over \sim \;$}}

\def\gcm3{{\rm\,g\,cm^{-3}}}
\def\ncm3{{\rm\,cm^{-3}}}

\def\>{$>$}
\def\<{$<$}

\received{}
\begin{document}

\begin{frontmatter}


\title{Exploring the High-Energy Cosmic Ray Spectrum with a Toy Model of Cosmic Ray Diffusion}

\author{Roger Clay \& Roland M. Crocker }
\address{School of Chemistry and Physics\\
University of Adelaide\\
5005, Australia\\
roger.clay,roland.crocker@adelaide.edu.au}

\date{\today}

\begin{abstract}

We introduce a static toy model of the cosmic ray (CR) universe in which cosmic ray propagation is taken to be
diffusive and cosmic ray sources
are distributed randomly with a density the same as that of local $L_*$ galaxies, $5 \times 10^{-3}$ Mpc$^{-3}$. 
These sources ``fire" at random
times through the history of the universe but with a set expectation time for the period between bursts.
Our toy model model captures much of the essential CR physics 
despite its simplicity and,
moreover, broadly reproduces CR phenomenology
for reasonable parameter values and without extreme fine-tuning. 
Using this model we investigate -- and find tenable -- the 
idea that the 
Milky Way may itself be a typical high-energy cosmic ray source. 
We also consider the possible phenomenological implications of the magnetic CR horizon
for the overall cosmic ray spectrum
observed at Earth.
Finally, 
we show that anisotropy studies should most profitably focus
on cosmic rays detected at energies above the so-called GZK cut-off, $\sim 6 \times 10^{19}$ eV.

\end{abstract}

\end{frontmatter}


\section{Introduction}

In the following study, we invoke a toy model of cosmic ray diffusion, in a static universe, 
 from a random ensemble
of extragalactic sources distributed throughout a cube with 2400 Mpc sides and the Earth at the center.
We assume a constant source density equal to $5 \times 10^{-3}$ Mpc$^{-3}$.
This is approximately
the local density of Milky-Way-like galaxies 
as determined in \cite{Aublin2006} on the basis of the ratio
of the local star formation rate density and the Galactic star formation rate
(also see \cite{Loeb2002}).
Each such source is taken to ``fire" at random times through its history, but with a
well-defined expectation for the time between such firings 
($t_{\rm wait} \equiv 1 $ Mpc/$c \simeq 5 \times 10^6$ year).

The power density of extragalactic cosmic ray sources can
be estimated by requiring it support the observed, high-energy spectrum
against losses.
On the basis of normalizing to the observed spectrum
at $10^{19}$ eV 
this has been determined to be around $5 \times 10^{44}$ erg yr$^{-1}$ Mpc$^{-3}$ \cite{Waxman1995,Blasi2004}.
With this input,
the typical energy into CRs from one of the outbursts we model must be $\sim 5 \times 10^{53}$ erg,
an amount of energy much too large to be associated with a single supernova but
only $1 - 10 \%$ of the energy released in 
a period of Seyfert activity or star-bursting in a ``typical" galaxy 
(see \cite{Veilleux2005} and \cite{Bland-Hawthorn2003} and references therein). 
We discuss the naturalness of these various scales further below.

Cosmic rays from our assumed extragalactic sources are taken to diffuse through
a purely turbulent magnetic field.
We assume that our modeled CRs are dumped {\it directly} into the extra-cluster space.
This allows the derivation, as a function of magnetic field amplitude and coherence length, 
of an upper limit (at any given energy) 
on the physical distance to the {\it magnetic horizon}
beyond which CRs could not have originated 
given the time available (given the age of the universe
and energy loss processes) to diffuse through the intervening fields.

We aim to sample field strengths representative of those found in 
extra-cluster space.
{\it Inside} clusters many lines of evidence now point to magnetic field amplitudes at the
few $\mu$G level out to distances of $\sim$ Mpc from cluster cores \cite{Clarke2001}, quite large enough
to affect the propagation of CRs to the highest energies.

Unfortunately, constraints on extra-cluster magnetic field strengths are not particularly strong.
At the super-cluster scale of $\lesssim$ 10 Mpc
there is evidence for magnetic field amplitudes
at the few $\times 0.1 \ \mu$G level \cite{Kim1989}.
For fields extending over cosmological distances
(i.e., on scales exceeding those
pertinent to
the filaments and walls of large scale structure,
$\sim$ 50 Mpc \cite{Widrow2002})
upper limits are in the range 1-10 nG for coherence lengths in the range
1-50 Mpc, on the basis of 
 examination of the rotation measures of
distant QSOs \cite{Blasi1999}.
Finally, we note that regular field components  -- neglected in our model --   
would tend to push out the magnetic horizon in certain directions (and pull it in elsewhere) 
given the phenomenon of CR drift in such fields.
The regular component is, however, expected to be very low 
in extragalactic space so that CR transport should be dominated, as we shall assume, 
by phenomena
associated with the turbulent field component \cite{Parizot2004}.

To sample, then, the reasonable parameter space 
we investigate field amplitudes of 1, 10 and 100 nG.
We also consider 
two coherence lengths:
10 kpc 
and 1 Mpc. 
These values bracket the range of scales from
galactic to cluster-size
and, therefore, the extragalactic magnetic field
coherence ``length" must fall within this range.
We neglect any possible evolution of all these quantities 
in our modelling.

Additionally, we model CR diffusion away from a single
local source -- of the same average power as an extragalactic source -- 
that is located at a distance equal to our separation from the center of the Galaxy, 
$\sim 8.5 $ kpc.  
The purpose of simulating the  additional, local source is to examine the consequences of the assumption
that the Milky Way (MW) be, itself, a typical CR source.
This idea, labeled the {\it holistic source model} by Aublin \cite{Aublin2006}, has
been examined by a number of authors (see, e.g., \cite{Milgrom1996,Loeb2002,Dermer2002,Aublin2006}).

Note that, if, by hypothesis, the MW is 
(or has contained) a typical CR source, 
then it must be capable of producing -- at least on occasion -- CRs at energies
up to and exceeding $10^{19}$ eV.
As a corollary to this statement, because CRs at such energies will not be greatly deflected
away from rectilinear propagation over Galactic length scales by the Galactic magnetic field,
the fact that there are no well-established anisotropies associated with Galactic structures
(Galactic plane, Galactic center) at these energies sets a minimum time scale to the time
since the last firing of the MW CR source cite{Giler1983,Giller2000}. 
Of course, this is not to say that we necessarily do not detect $\gtrsim 10^{19}$ eV CRs from past
firings of the putative MW source, only that any we do detect have been diffusion processed to
the extent that they are almost equally likely to come from any region of the sky. 

We do
not know the properties of the {\it local} magnetic field  
(immediately outside our
own Galaxy) but we assume that it is {\it unlikely} to have a field
strength as low as is found at a great distance from any galaxy
cluster.  
We asume, therefore, 
that the magnetic field in the nearby interior of the cluster
within which our Galaxy is located is at or above
0.1 $\mu G$. 
We chose to sample field strengths of 0.1,0.2, and 10 $\mu {\textrm G}$ in our modelling,
the latter representing an extreme value (a direct extension of that typical for the
Galactic plane field strength). 
The local field is assumed to have a spherically-symmetric structure
and a coherence length of 10 kpc.
We emphasise that the local field introduced here
does {\it not} represent the field typical for the Galactic disk (which is known to have
both turbulent and regular components and a typical total amplitude of few $\mu{\textrm G}$ near the Earth), 
but rather the larger, cluster-scale magnetic field within which the
entire Galaxy is situated.
The effect of confinement in this latter field can be
calculated by following the procedure described by \cite{Clay2002}
but is neglected here as it does not affect the
spectrum at energies of $10^{18}$ and above with which we concern ourselves.

In summary our physical picture is 
one in which at energies 
above the spectral upturn in the cosmic ray spectrum at $\sim 3 \times 10^{18}$ eV (the {\it ankle})
there is overlap between the CR diffusion spheres around individual galaxies 
and we measure a flux of CRs that is the same as anywhere else (including inter-galactic space)\cite{Aublin2006}, 
i.e., universal.
In contrast, at considerably lower energies we measure at earth a flux of CRs that
originates from the Milky Way and
is over-abundant with respect to the universe-at-large. 
At the lower end of the region of concern to us ($\sim 10^{17}$ eV), 
the Galactic CR spectrum is in approximate steady-state in our model because
the diffusion time from the Galactic center to us through the local field is longer than
the expected time between firings of this local source. 
At the upper end of the Galactic spectrum
(but below the region where the extragalactic flux becomes dominant)
this condition is not satisfied, however, and
the spectrum is time-dependent. One must tune to the time of
the last Galactic CR outburst to arrive at a  spectrum
consistent overall with observations. 



Our picture, then,
is conventional in the sense that the
ankle
is associated with the transition from dominance by the Galactic source
to dominance by the extragalactic sources. 
We show below, however, that the assumption that 
the Milky Way be a typical CR source ameliorates a fine-tuning problem
implicit in this interpretation of this structure, viz.
{\it why is the ankle placed where it is such that we can observe an extragalactic flux but this flux does not
dominate the Galactic CR flux?}

Finally, we note in passing here
that alternative recent models (see \cite{Berezinsky2007} and references therein)
that would posit a transition at considerably
lower energies, $\sim 10^{17.5-18}$ eV, and explain the ankle as the result of 
a dip in the spectrum of extragalactic protons because of their
Bethe-Heitler pair production collisions on the CMB, would seem
to suffer an even more extreme fine-tuning problem.
This, namely, is the matching  between the normalizations
of the Galactic and extragalactic components
required so that the transition 
is indicated by no strong spectral feature at all or,
at least, one as weak as the spectral downturn represented by the
so-called second knee.

\section{Diffusive Transport}

In this work we assume that CR transport can be described as a purely diffusive process.
Following, e.g., \cite{Gaisser1990}, 
denoting the density of protons at position $\mathbf{x}$ and with energy between
$E_p$ and $E_p + d E_p$ and at time $t$ by $N_p(E_p,\mathbf{x},t)$, 
the proton transport equation with acceleration, 
convection, and collision losses and gains all neglected can be written:
\begin{equation}
{\dot N} \, = \, \nabla \cdot (D \nabla N) \, + \, Q \, ,
\label{eqn_transport}
\end{equation}
where $Q$ is an explict source of particles and $D$ is the diffusion coefficient which, formally, relates the current of particles
(in our case protons) to a spatial gradient in the density of such particles \cite{Gaisser1990}.

The Green's function for Eq.(\ref{eqn_transport}) -- 
which, physically, gives the probability for finding a particle, that was injected
at the origin, at a position $\mathbf{r}$ after a time $t$ -- is
\begin{equation}
G(\mathbf{r},t) \, = \, \frac{1}{8 (\pi \, D \, t)^{3/2}} \exp\left( \frac{-\mathbf{r}^2}{4 \, D \, t} \right) \, .
\label{eqn_greens}
\end{equation}
We use Eq(\ref{eqn_greens}) to determine the CR flux due to each CR outbursts, at some given time previous to now,
from each simulated CR source at (random) position $\mathbf{r}_{\rm source}$. 

The lower limit on the diffusion coefficient is given by the Bohm case wherein
the scattering distance is equal to the gyroradius implying that
\be
D_{\rm Bohm}(E_p, B) \, = \, \frac{c \, r_{\rm gyro}}{3} \, ,
\ee
where the gyro-radius is, in general given by
\begin{equation}
r_{\rm gyro}(p,Z,B) = \frac{p \, c}{Z \, B} 
\simeq {\rm 1 pc} \, \left( \frac{E}{\rm PeV} \right) 
\, \left( \frac{Z \, B}{\rm \mu G} \right)^{-1}\, ,
\label{eqn_gyro}
\end{equation}
(where $Z$ is the CR's charge in units of the charge on the proton and $p$ its 
momentum\footnote{Note that in this work we only model the propagation of protons so that
$Z \ = \ 1$ in equation \ref{eqn_gyro} and throughout.}).

A more realistic behavior for the diffusion coefficient -- that we employ in our modeling -- is as
parameterized on the basis of the numerical work of Parizot
\cite{Parizot2004} who considers particle transport in a purely
turbulent field with a Kolmogorov spectrum:
\be
D(E_p) \, = D_\star \left[\left(\frac{E}{E_\star}\right)^{\frac{1}{3}} 
+ \left(\frac{E}{E_\star}\right) + 
\left(\frac{E}{E_\star}\right)^2\right]\, .
\ee
where $E_\star$ is implicitly defined via
\be
r_{\rm gyro} (E_\star) \equiv \frac{\lambda_{\rm coh}}{5},
\ee
and
\be
D_\star \equiv \frac{1}{4} c r_{\rm gyro} (E_\star) \, .
\ee
In these equations $\lambda_{\rm coh}$ denotes the coherence length of the magnetic field.

As implicit in the above equations,
the investigations of \cite{Parizot2004} reveal that for particle energies such that the gyro-radius
considerably exceeds the coherence length of the magnetic field, the scattering length
scales as $E^2$. 
At the other extreme, for low energies such that $r_{\rm gyro} \ll \lambda_{\rm coh}$ and for a 
Komogorov spectrum of magnetic field turbulence, the scattering length scales as $E^{1/3}$.
Finally, when $r_{\rm gyro} \sim \lambda_{\rm coh}$ a Bohm-like scaling is approximately followed
for a limited energy range
with the scattering length $\propto E$ (though note the coefficient of proportionality never
actually falls to the Bohm value: $D \gtrsim 3 D_{\rm Bohm}$; \cite{Parizot2004}).

\section{Energy Loss in Extra-Galactic Space}

In propagating over cosmological distances, cosmic ray protons lose energy through red-shifting and via inelastic collisions
with background light fields (through both Beth-Heitler pair-production and resonant photo-pion production),
the cosmic microwave background most significantly at energies $\gtrsim 10^{19}$ eV 
\cite{Greisen1966,Zatsepin1966}. 
We account for the modification induced by these effects on input spectra via a
parameterization of the energy-dependent attenuation length
calculated by \cite{Geddes1996} and presented in their fig.1.
Note that we neglect protons down-shifted in energy through their interactions,
an approximation that works reasonably for power-law spectra $\propto E^{-2}$
or steeper such as we investigate here.

\section{Results}

\begin{figure*}[ht]
\epsfig{file=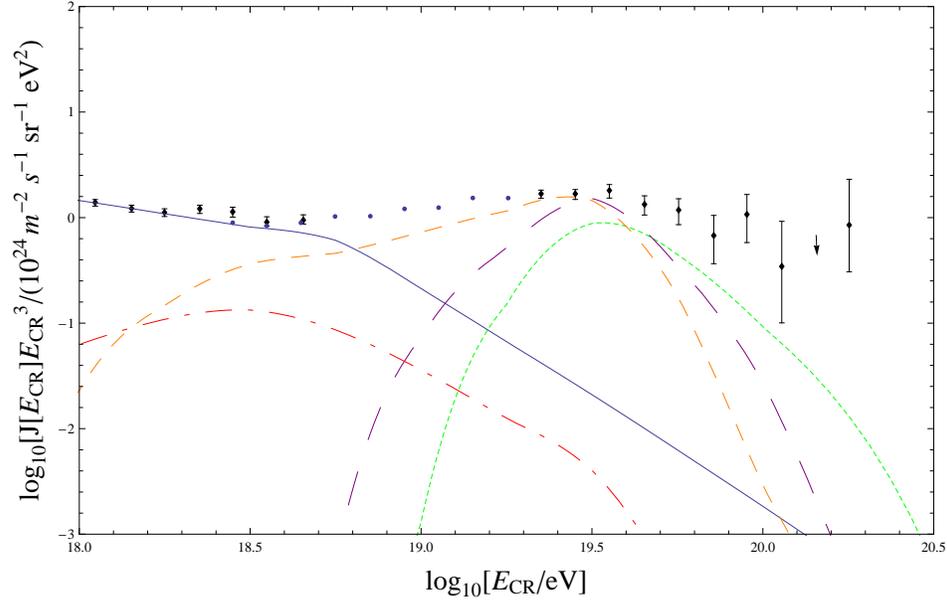,height=8cm,angle=0}
\caption{Energy-cubed weighted fluxes for 
a local source in a local field of 200 nG with last outburst 6 Myr ago
(shown as the solid (blue) curve)
plotted against the following extragalactic cases (all calculated assuming
a 10 kpc coherence length for the extragalactic magnetic field structure):
dotted (green) -- 100 nG intergalactic field;
 long dash (purple) -- 30 nG field;
short dash (orange) -- 10 nG;
dot-dash (red) -- 1 nG;
data points -- Auger spectrum with statistical errors \cite{VanElewyck2007}
}
\label{fig_plot1.eps}
\end{figure*}

We plot some representative results of our spectral modeling in
figure \ref{fig_plot1.eps}.  
In this figure, we show
spectra derived from a randomly-placed ensemble of
extragalactic proton sources.  
The cosmic
rays have propagated through turbulent magnetic fields (all with a coherence length of 10 kpc) and
of
strength 1, 10, 30, and 100 nG and had $E^{-2}$ source spectra at injection.
Diffusive propagation is a slow process and the magnetic horizon
is increasingly important towards lower particle energy  and in the case
of strong magnetic fields.  This effect is evident in the
figure. We note that for the strongest magnetic field pictured there is a complicated
interaction between the GZK and magnetic horizon effects which results in not only
the expected systematic shift of the peak of the $E^3$-weighted spectrum to the right but also an
overall attenuation of this weighted spectrum.

The effect of changing the magnetic field coherence length 
can, again in concert with the GZK attenuation,
 also result in complicated phenomenology as shown in figure 2.
We noted above that the ratio of the gyroradius to the coherence length determines the scattering properties of 
particles in the turbulent field. For a 10 nG field strength the proton gyroradius is $\sim$Mpc
at $10^{19}$ eV and we therefore expect an appreciable dependence of the spectrum on the assumed 
coherence length at these energies.

The Milky Way galaxy is of unique importance as a CR source but, as presaged above, in our
model it can be added in the same way as the random galaxies.
This is also shown in the figure, although we now have to
explicitly specify the time of the most recent outburst as, in
this case, that outburst can have a dominating effect.  
The figure shows results for a magnetic field
of 0.2 $\mu{\textrm G}$
with the most recent local outburst six million years ago -- a case which reproduces the data tolerably well.
We have found that, as expected,  ``dialing-up" the local field strength or
taking a more recent time for the last local outburst
increases the amplitude of the 
local component relative to the extragalactic contribution.

Note that our procedure, using only
assumptions of similarity between all galaxies such as the Milky
Way and a common mean time between identical outbursts of three
million years, provides a selection of spectra which are not
dissimilar to those observed and commonly assumed to be Galactic and
extragalactic. Apart from selecting within the modest range of
plausible field parameters shown in the figure, there is no
arbitrary normalisation between the spectral components.  

If we select results for a 10 nG intergalactic with 1 Mpc coherence length
field and a 0.2 $\mu \textrm{G}$ local field,
we get a combined
spectrum plausibly similar to that which is observed
especially considering the scale of the uncertainties introduced by
the imprecisely-known input parameters (particularly, the source density and expectation
time between firings) and the admitted crudity of our model
(which assumes all sources have the same time-integrated power, a single coherence
length for the fields rather than a distribution, etc).
Figure 3
shows the summed spectrum due to these components
and 
the  spectrum measured by Auger
\cite{VanElewyck2007}.
We also emphasise that the error bars on the experimental
data only show statistical errors, not systematic, which, in reality
can be considerable. 
The Auger collaboration \cite{Parizot2007} states that, at present,
the statistical and systematic uncertainties within the relevant
energy scale are 6\% and 22\% respectively.

Also plotted in Figure 3 is the extragalactic flux expected 
for the same 10 nG field but with the 
opposite extremum of coherence length, viz. 10 kpc.
In such a field structure
it can be ascertained that CR diffusion becomes too fast
at energies in the vicinity of the GZK ``cut-off" energy for the highest 
flux to be reproduced. The spectrum, therefore, dies away too quickly at high energy.
We emphasise that this cut-off in the spectrum in not 
simply given by the GZK effect 

In general, one can see that in order to reproduce the observed spectrum a dip in the range
18.5 $\gtrsim$ log($E_{CR}$/eV) $\gtrsim$ 19.0 is required. 
This sets a constraint on the extragalactic field strength in the range 10-100 nG, at least within
our toy model.
For field intensities $\gtrsim 100$ nG the dip in the overall (extragalactic + Galactic sources)
is too large.
At field intensities $< 10$ nG there is insufficient flux above $10^{19}$ eV 
to reproduce the observed spectrum.

Again, at least within our toy model, we can constrain the time
of the last outburst of the MW source.
If, as we assume, there is a 
halo magnetic field local to the intra-cluster space around the Galaxy 
then, for a local field intensity of 10 $\mu$G, the last outburst
was earlier than 30-300 million years ago (lest the MW source signal
completely swamp all other sources 
meaning that the observed dip in the spectrum cannot be reproduced).
Of course, this is much longer than the expected time between outbursts from the MW
if it is an ordinary CR source and, therefore, represents an extreme fine-tuning indicating that
this choice of local field intensity is disfavored.
For a 1 $\mu$G field, the last outburst is required to be more than 30 million years ago, 
again a fine-tuning given the 3 million year expectation time.
Finally, for a 0.1 $\mu$G field, the last outburst should be in the range 1-10 million years
and we find that, at least from the point of view of naturalness, 
that this range of amplitude would seem to be favored by our model
(we chose a 0.2 $\mu$G field).

\subsection{Anisotropy}

In the diffusive regime, we can investigate the relative
anisotropy due to each of the assumed sources.  
Our analysis employs
simple diffusion ideas as expressed, for example, in \cite{Allan1972}, 
in which the anisotropy is given in magnitude by the
ratio of the scattering mean free path (or gyroradius under some
circumstances) to the source distance.  Our results for the
spectrum model just described give a low anisotropy of below 1\%
between 0.2 EeV and 30 EeV.  However, at energies at which the
GZK cut-off is important, the anisotropy then rises rapidly with
energy and is above 10\% when 100 EeV is reached.  The detail of
this energy dependence will clearly depend on the real local
distribution of galaxies.  However, the important, and perhaps
not surprising, point is that the fact that the GZK effect limits
the distance to observable sources makes the actual spatial
distribution of sources observable.  In a conventional model for
AGN sources of the highest energy cosmic rays, this result is
obviously intuitive, but that is not so for our picture of bursts
from conventional galaxies.

\section{Discussion -- Naturalness of Inferred Scales}

Bland-Hawthorn and Cohen \cite{Bland-Hawthorn2003} have, on the basis of 
infra-red, radio, and X-ray
observations, inferred the 
existence of a large-scale, bipolar wind out of the Galactic center.
This wind and other structures seen on large scales -- in particular, the North Polar Spur, 
an X-ray/radio loop that extends from the Galactic plane all the way to 
$b = +80^\circ$ \cite{Sofue2000} -- support, in turn, the notion
that the Galactic center \cite{Melia2001} is host to explosive outbursts 
of total energy $\sim 10^{55}$ erg that
occur every $\sim 10$ million years or so (see \cite{Bland-Hawthorn2003}
and references therein).
Sanders \cite{Sanders1981} long ago predicted a similar time scale and energetics for
intermittent activity of the Galactic center and similar spiral-galaxy nuclei
with a period of Seyfert luminosity ($L > 10^{43}$ erg s$^{-1}$) of duration $\sim 10^5$ years expected
every $\sim 10^7$ years.
In Sanders' picture this activity is driven by
the intermittent accretion of gas in giant molecular clouds on to the galaxy's central
black hole. 
In contrast, Bland-Hawthorn and Cohen prefer an interpretation of their observations in terms of star-bursting. 
In fact, activity of the central black hole and nuclear star bursting may be closely inter-related. 
For instance, Nayakshin and Cuadra \cite{Nayakshin2005} favor a picture in which, some millions of years ago,
our Galaxy was ``robbed" of the chance for truly bright AGN activity because
the gaseous fuel that would otherwise have been available to power this activity was 
driven away by a nuclear star-burst. 
Nevertheless, in their picture, the MW achieves a Seyfert luminosity.

Regardless of the particulars, similar
energetics and timescales for periodic Galactic center activity emerge from 
all of the above and, assuming $\sim 5\%$
of the total energy released in each event ends up in high-energy protons, the
energetics and outbursting timescales we infer
for the putative local CR source match these scales nicely.
In this  connection, we note that
the scale of our inferred Galactic center CR ``explosions"
are well inside the contraints on CR flux implied by the non-detection
of Li I and B I in the Sgr A molecular cloud \cite{Lubowich1998}.

It is also interesting to note that Giller and co-workers \cite{Giler1983,Giller2000}
have long advocated the idea that CR outbursts from the Galactic center 
may account for much of the observed, high energy CR spectrum.
Further supporting our general picture and our favored parameters,
Giller's work has tended to support 
a periodicity of up to 10 million years for the GC outbursts in order that anisotropy
upper limits are obeyed.

Another interesting result is that the energy density in the favored field strength 
for the local (halo) field, 0.2 $\mu$G, is $\sim \times 10^{-3}$ eV cm$^{-3}$
which is close to the energy density represented by the 
cosmic ray spectrum above the knee in the spectrum.

\section{Conclusion}

We have demonstrated a toy model that can reproduce the broad phenomenology of the 
ultra-high energy cosmic ray spectrum for reasonable parameter values. 
Our model invokes a random distribution of extragalactic sources 
with a space density equal to that observed in the local universe for Milky Way-like 
galaxies that ``fire", i.e. explosively inject, an $E^{-2}$ spectrum of cosmic
ray protons every three million years on average. 
By construction, our model also incorporates a local source located at a distance of 8.5 kpc,
the approximate distance to the Galactic center.
From our modeling, we favor a local field (but external to the Milky Way disk) of
an amplitude in the $\sim$ 0.1 $\mu$G range with the most recent firing of the 
Milky Way source to be at $\sim 6$ million years, a comfortable multiple  
(2) of the average temporal separation between firings of the average source 
(which we assume the local source to be).
As far as the extragalactic contribution to the spectrum goes, our modeling favors an 
average intergalactic field of 10 nG amplitude and an average coherence length for this field
toward the longer end of the allowable range, viz. 1 Mpc.

We note that these parameter choices are probably not unique 
nor do they perfectly reproduce the observed 
spectrum. 
Given the crudity of our model, however, and the uncertainty in the input parameters,
we think we have demonstrated the tenability of the broad ideas our model instantiates

A more detailed model than that under consideration would, in particular, take into account the known distribution of
local galaxies and might, in addition, assign a variable cosmic ray-power
to each of these known galaxies correlated with, say, the inferred mass of the central black
hole or with some other correlate of past cosmic ray activity.
Such considerations
are potentially
of particular relevance in the case of Andromeda, the closest large galaxy to the
Milky Way.
This object is located at a distance of only $\sim$ 800 kpc, significantly closer than the expected distance
to the closest Galaxy in our model, viz. 
$\sim (5 \times 10^{-3} \, \textrm{Mpc}^{-3})^{-1/3} \simeq 6$ Mpc.
It also hosts a supermassive black hole of $\sim 10^8 \solarmass$ \cite{Bender2005}, substantially
larger than the $\sim 4 \times 10^6 \solarmass$ black hole located at the Galactic center
and may, therefore, have been a significantly more powerful CR emitter that the MW 
in its history.
Such considerations will be addressed in a future work.

\begin{figure*}[ht]
\epsfig{file=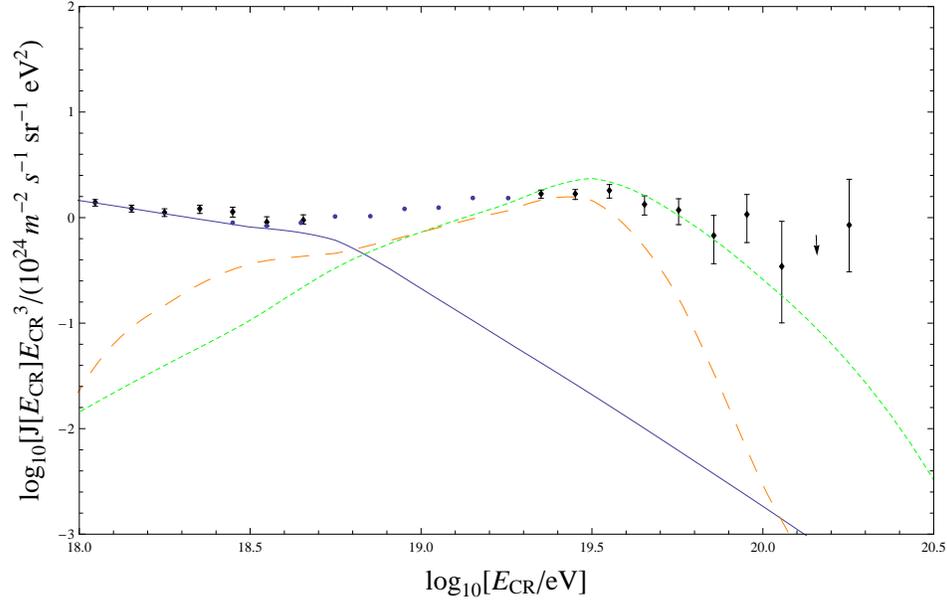,height=8cm,angle=0}
\caption{Energy-cubed weighted fluxes for 
a local source in a local field of 200 nG with last outburst 6 Myr ago
(shown as the solid (blue) curve)
plotted against the following extragalactic cases (all calculated assuming
a 10 nG amplitude for the extragalactic magnetic field structure):
short dash (orange) -- 10 kpc coherence length;
dotted (green) -- 1 Mpc coherence length;
data points -- Auger spectrum with statistical errors \cite{VanElewyck2007}
}
\label{fig_plot2.eps}
\end{figure*}

\begin{figure*}[ht]
\epsfig{file=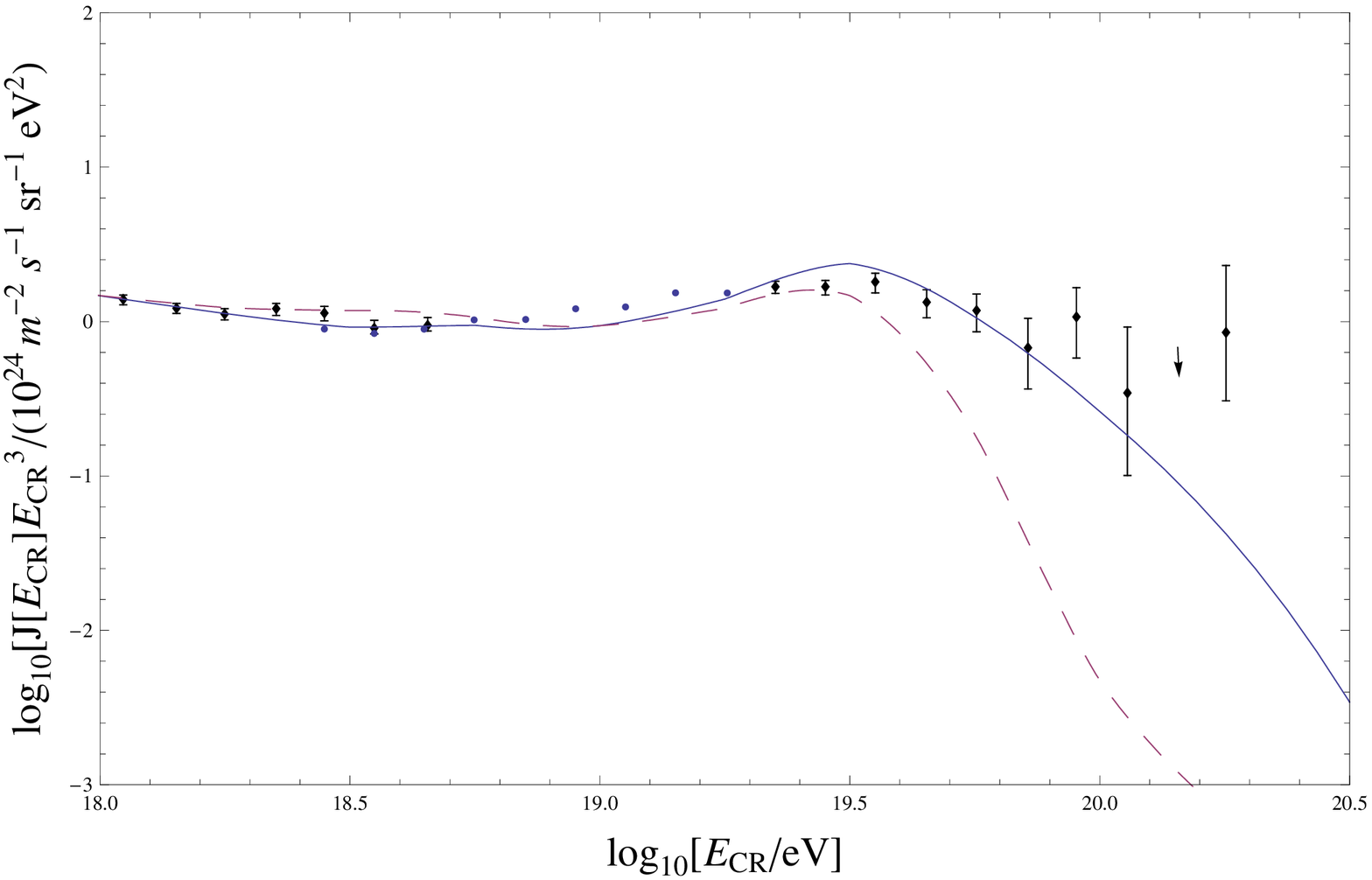,height=8cm,angle=0}
\caption{Modeled Galactic + extragalactic fluxes
weighted by energy-cubed.
In both cases the Galactic spectrum is calculated assuming
a 0.2 $\mu$G local field and a most-recent firing 6 Myrs ago.
The extragalactic fluxes are for the following cases:
solid (blue) -- 10 nG intergalactic field with 1 Mpc coherence length;
long-dash (purple) -- 10 nG intergalactic field with 10 kpc coherence length;
data points -- Auger spectrum with statistical errors \cite{VanElewyck2007}.
}
\label{fig_plot3.eps}
\end{figure*}

\section{Acknowledgements}
The authors acknowledge enlightening conversations Todor Stanev.
RMC is supported at the University of Adelaide by Ray Protheroe
and Ron Ekers'
Australian Research Council's Discovery funding scheme grant
(project number
DP0559991).

\end{document}